\documentclass[11pt,a4paper]{article}
\usepackage{amsfonts}
\def\be{\begin{equation}}
\def\ee{\end{equation}}
\def\bea{\begin{eqnarray}}
\def\eea{\end{eqnarray}}

\begin{document}
\setcounter{footnote}{1}
\renewcommand{\thefootnote}{\fnsymbol{footnote}}
\noindent
{\Large \bf Is Physics Asking for a New Kinematics?}
\vskip 0.7cm
\noindent
{\bf R. Aldrovandi\footnote{E-mail: ra@ift.unesp.br} and J. G. Pereira\footnote{E-mail: jpereira@ift.unesp.br}}\\ 
{\it Instituto de F\'{\i}sica Te\'orica, Universidade Estadual Paulista \\
Rua Pamplona 145}, {\it 01405-900 S\~ao Paulo, Brazil}

\vskip 0.8cm
\noindent
{\bf Abstract~}{\footnotesize It is discussed whether some of the consistency problems of present--day physics could be solved by replacing special relativity, whose underlying kinematics is ruled by the Poincar\'e group, by de Sitter relativity, with underlying kinematics ruled by the de Sitter group. In contrast to ordinary special relativity, which seems to fail at the Planck scale, this new relativity is ``universal'' in the sense that it holds at all energy scales.}

\section{Introduction}
\setcounter{footnote}{0}
\renewcommand{\thefootnote}{\arabic{footnote}}

The first kinematic group in physics was the Galilei group, under which Newtonian classical mechanics is invariant. This invariance has later received the name of ``Galilean relativity''. By the end of the nineteenth century, inconsistencies between Newtonian mechanics and electromagnetism have triggered the search for another group, another relativity. That search culminated in the establishment of special relativity, whose underlying kinematics is ruled by the Poincar\'e group ${\mathcal P}$. Since then, this theory has been the fundamental paradigm underlying all relativistic theories.

One century later, physics is again facing intricate consistency problems. Two examples are (i) the apparent inconsistency of general relativity with quantum mechanics, known as the quantum gravity problem, and (ii) the acceleration in the universe expansion rate, known as the dark energy problem. Would these problems mean that we need a new relativity? The question is not new~---~there are theoretical arguments suggesting that the Poincar\'e symmetry might break down at ultra--high energies. The basic point is the existence, at the Planck scale, of an invariant length parameter,  the Planck length. Since a length contracts under a Lorentz boost, the Lorentz symmetry is proposed to be broken at that scale~\cite{lorentzX}. Relying on this argument, many attempts have been made to find such a new relativity, most of them based on the $\kappa$--deformed Poincar\'e group.\footnote{The relevant literature can be traced back from the papers cited in Ref.~\cite{dsr}.}

The largest symmetry group on a 4-dimensional spacetime is the conformal group, of which ${\mathcal P}$ is a subgroup. From the kinematic point of view, Poincar\'e relativity can be viewed as describing the implications to Galilei relativity of introducing an invariant velocity scale~---~the speed of light $c$~---~into the Galilei group. Conversely, Galilei relativity can be obtained from Poincar\'e's by taking the formal limit of the velocity scale going to infinity (non-relativistic limit). The algebraic hierarchy between these two groups is founded on the Wigner--In\"on\"u process of group contraction and expansion~\cite{inonu}. Within this point of view, and taking into consideration the existence of an invariant length parameter at the Planck scale, it is natural to expect that an ultra--high energy kinematics would emerge from introducing {\em both} a velocity {\em and} a length scales into the Galilei group. The natural candidate is then the de Sitter group, another subgroup of the conformal group. While preserving a certain special length--parameter $l$, which is related to the cosmological term $\Lambda$ through
\be
\Lambda = \frac{3}{l^2},
\label{pureds}
\ee
it presents the same kind of algebraic hierarchy described above.

In the formal limit of a vanishing cosmological term ($l \to \infty$), the de Sitter group contracts to Poincar\'e~\cite{gursey}, in which only the velocity scale $c$ is present. A further limit $c \to \infty$ leads Poincar\'e to Galilei relativity. It is interesting to observe that the order of the group expansions (or contractions) is not important. If we introduce in the Galilei group an invariant length parameter, we end up with the Newton-Hooke group~\cite{nh}, which describes a (Galilean) relativity in the presence of a cosmological constant. Adding to this group a fundamental velocity scale, we end up again with the de Sitter group, whose underlying relativity involves both a velocity and a length scales. Conversely, the low--velocity limit of the de Sitter group yields the Newton-Hooke group, which contracts to the Galilei group in the limit of a vanishing cosmological constant.

Now, replacing Poincar\'e by the de Sitter group means generalizing ordinary special relativity to a de Sitter relativity~\cite{dssr}. This, in turn, means that any physical system must modify the local structure of spacetime in such a way that {\it the region occupied by the system becomes a de Sitter spacetime.}\footnote{This hypothesis has already been proposed by F. Mansouri in a different context~\cite{mansouri}.} In addition to the usual gravitational field, therefore, any physical system must engender a further local de Sitter field whose intensity --- measured by the local value of $\Lambda$ --- is proportional to its energy density. The natural question then arises: how does a physical system give rise to such a field? This question, which belongs to the realm of de Sitter relativity, is the main issue of the next section.

\section{Fundamentals of de Sitter Relativity}
\label{sec:dsr}

\subsection{The de Sitter Space and Group}

A de Sitter spacetime, which will be denoted $dS(4,1)$, is a homogeneous space defined as the quotient between de Sitter and Lorentz groups~\cite{HE}:
\be
dS(4,1) = {SO(4,1)}/{\mathcal L}.
\ee
Immersed in a five--dimensional pseudo--Euclidian space ${\bf E}^{4,1}$ with Cartesian coordinates\footnote{We use the capital Latin alphabet $(A, B, C, \dots = 0,1,2,3,4)$ to denote indices related to the pseudo--Euclidian ambient space coordinates, and the lowercase Latin alphabet $(a,b,c, \dots = 0,1,2,3)$ to denote de Sitter algebraic indices, which are raised and lowered with the Lorentz metric $\eta_{ab} = \mbox{diag}~(+1,-1,-1,-1)$.} $(\chi^A) = (\chi^a, \chi^{4})$, it is defined by
\be
\eta_{a b} \, \chi^{a} \chi^{b} + (\chi^{4})^2 
= - \,\, l^2,
\label{dspace1}
\ee
with $l$ the de Sitter length--parameter. Notice that, though we are talking about {\em the} de Sitter space and group, it should be clear that there are actually infinite such spaces and groups, one pair for each value of $l$ (or $\Lambda$).

In stereographic coordinates $\{x^a\}$~\cite{gursey}, the generators of the de Sitter Lie algebra are written as
\be
L_{ab} = \eta_{ac} \, x^c \, P_b - \eta_{bc} \, x^c \, P_a,
\ee
and
\be
L_{a4} = l P_a - (4l)^{-1} K_a,
\label{dstransitivity}
\ee
where
\be
P_a = \partial_a \quad \mbox{and} \quad
K_a = \left(2 \eta_{ac} \, x^c x^b - \sigma^2 \, \delta_a^b \right) \partial_b
\ee
are, respectively, the generators of translations and proper conformal transformations. Generators $L_{ab}$ refer to the Lorentz subgroup, whereas the remaining $L_{a4}$ define the transitivity on  de Sitter spacetime. To make contact with the Poincar\'e group, it is convenient to define the generators
\be
\pi_a \equiv \frac{L_{a4}}{l} = P_a - (4l)^{-2} K_a,
\ee
which are usually called de Sitter ``translation'' generators~\cite{livro}.

From the algebraic point of view, the change from Poincar\'e to de Sitter is achieved by replacing $P_a$ by $\pi_a$. As a consequence, the conformal transformations will naturally be incorporated in the spacetime kinematics. The relative importance of translations and proper conformal transformations, as can be seen from Eq.~(\ref{dstransitivity}), is determined by the value of $l$, that is, by the value of the cosmological term. It is also important to note that, since the de Sitter group involves an invariant length--parameter~---~in addition to the speed of light~---~de Sitter special relativity can be interpreted as a kind of {\em doubly special relativity}~\cite{dsr}. There is a crucial difference, though: whereas in the usual models of doubly special relativity the Lorentz symmetry is assumed to be violated, in de Sitter special relativity only translations are violated, the Lorentz subgroup remainig as a physical symmetry~\cite{lorentz}.

\subsection{Horizons and Fundamental Constants}
\label{horizon}

In {\em static} coordinates $(t, r, \theta, \phi)$, the de Sitter metric is written as
\be
ds^2 = (1- {r^2}/{l^2} ) c^2 dt^2 - \frac{dr^2}{\left(1-r^2/l^2 \right)} -
r^2 (d\theta^2 + \sin^2\theta d\phi^2).
\label{staticor}
\ee
In this form, it reveals an important property of the de Sitter spacetime: the existence of a horizon at $r = l$. Now, as is well known from the Schwarz\-schild solution, there is a remarkable relation between gravitational horizons and thermodynamic properties~\cite{bhth}. The common presence of a horizon makes it possible to  attribute thermodynamic features  to the de Sitter horizon \cite{gh} in the same way as in the Schwarz\-schild case. This result can be demonstrated in many different ways,  the simplest one being probably that based on the relationship between temperature and the Euclidian extension of spacetime. Spacetimes with horizons present a natural analytic continuation from Lorentzian to Euclidean signature, obtained by making $t \to i t$. If the metric becomes periodic, one can naturally associate a notion of temperature to such horizons, and consequently any other thermodynamic quantity. For example, one can associate to the de Sitter horizon the entropy
\be
S_{dS} = \frac{k_B \, A_h}{4 l_P^2 \gamma},
\label{dsentro}
\ee
where $k_B$ is the Boltzmann constant, $A_h = 4 \pi l^2$ is the area of the horizon, $l_P = \sqrt{G \hbar/ c^3}$ is the Planck length, and $\gamma$ is a parameter which can be interpreted as a de Sitter version of the Barbero-Immirzi parameter~\cite{BaIm}.

On the other hand, we know from quantum mechanics that there is a lower limit for all physical quantities. For example, the smallest amount of an electromagnetic field, a photon, is determined by the Planck constant as a quantum of the field. In a similar fashion, the smallest possible length is the Planck length. Since in de Sitter relativity there is a free length parameter $l$, it is natural to assume that its minimum value is the Planck length $l_P$. Relying upon this hypothesis, we can say that the quantum of entropy, that is, the smallest possible amount of entropy, is that associated with a de Sitter horizon of radius $l=l_P$. In this case, if we set $\gamma = \pi$, Eq.~(\ref{dsentro}) yields
\be
S_{dS} = k_B,
\ee
from which we see that the Boltzmann constant appears as a quantum of entropy. The entropy proportional to a spherical surface, like that of a soap bubble~\cite{Ise92}, cannot tend to zero: no sphere can be reduced continuously to a point.

Let us consider now the Boltzmann equation
\be
S_{dS} = k_B \ln \Omega,
\label{Boltz}
\ee
with $\Omega$ the number of states. Considering that we know the entropy, we can use it to express the number of states, which reads 
\be
\Omega = \exp \left( \frac{S_{dS}}{k_B} \right) = \exp \left( \frac{l^2}{l^2_P} \right),
\ee
where we have already used Eq.~(\ref{dsentro}). We see from this expression that, contrary to our common sense, the number of states is not given by the area of the horizon divided by the Planck area, but by the exponential of this number. For the minimum value $l = l_P$, it yields the minimum number of quantum states 
\be
\Omega = e, 
\ee
with the corresponding minimum entropy given by $S_{dS} = k_B$. This should be compared with the classical result, according to which $l$ can reach zero, the minimum number of states is consequently $\Omega = 1$, and entropy is allowed to vanish. 

\subsection{Ordinary Matter and the Cosmological Term}

We are now ready to answer the question posed at the end of section 1: how does a physical system give rise to a local $\Lambda$? Inspired in the previous discussion, let us consider first a de Sitter spacetime with $l = l_P$. The corresponding cosmological term is
\be
\Lambda_P = \frac{3}{l^2_P}.
\label{L1}
\ee
Considering that a cosmological term represents ultimately an energy density, we define the Planck   energy density 
\be
\varepsilon_P = \frac{m_P \, c^2}{(4 \pi/3) l_P^3} \, ,
\ee
with $m_P = \sqrt{c \hbar/G}$ the Planck mass. In terms of $\varepsilon_P$, Eq.~(\ref{L1}) assumes the form 
\be
\Lambda_P = \frac{4 \pi G}{c^4} \, \varepsilon_P  \, .
\label{kineLambda0}
\ee
Now, similarly to the entropy, whose quantum naturally emerges at the Planck scale, the very definition of $\Lambda_P$ can be considered a particular, extremal case of a general expression relating the energy density of a physical system to its corresponding ``cosmological'' term. Accordingly, to a physical system of energy density $\varepsilon$  will be associated 
\be
\Lambda = \frac{4 \pi G}{c^4} \, \varepsilon.
\label{kineLambda}
\ee
This equation gives the local value of the ``cosmological'' term as a function of the energy density of the physical system. It is important to reinforce that the $\varepsilon$ appearing in this equation is not the dark energy density, but the matter energy density. For small energy densities, $\Lambda$ will be very small, spacetime will approach Minkowski spacetime, and de Sitter special relativity will approach ordinary special relativity, whose kinematics is governed by the Poincar\'e group.

\section{Some Physical Consequences of de Sitter Relativity}
\label{sec:dsr2}

\subsection{de Sitter Kinematics and the Cosmological Constant}

As a first application of de Sitter relativity, consider the case in which the physical system is the whole universe. We take for $\varepsilon$ the Friedman critical energy density for the case of a spacetime with flat space section ($k=0$), which is given by
\be
\varepsilon = \frac{3 H_0^2 c^2}{8 \pi G} \, ,
\ee
with $H_0$ the Hubble constant. In this case, by  Eq.~(\ref{kineLambda}),  
\be
\Lambda \simeq \frac{3 H_0^2}{2 c^2} \,.
\label{coscon2}
\ee
If we write $H_0 = 100 \, h$~(Km/s)/Mpc, the cosmological term is found to be
\be
\Lambda \simeq 1.7 \, h^2 \times 10^{-56}~\mbox{cm}^{-2},
\ee
which is of the order of magnitude of the observed value~\cite{obs}. This simple estimate is given to illustrate the main point:  replacing Poincar\'e by de Sitter as the group governing the spacetime kinematics leads to a relation between the energy density of any physical system and the local value of $\Lambda$.  When applied to the whole universe, that relation is able to predict the value of the cosmological constant. The latter is no more an independent parameter~---~it is determined by the spacetime kinematics and is, in principle, calculable. 
  
\subsection{Speculations About Quantum Gravity}

According to quantum gravity arguments, at ultra-high energies the spacetime texture is expected to experience deep changes. According to de Sitter relativity, these changes have a very precise form: spacetime departs from Minkowski and becomes a de Sitter spacetime. As a consequence, de Sitter relativity naturally incorporates the conformal transformations in spacetime kinematics, which is directly related to the presence of $\Lambda$. For small energies, $\Lambda$ will be small and the local de Sitter spacetime approaches Minkowski, which is transitive under ordinary translations only. Near the Planck scale, $\Lambda$ will approach the Planck value (\ref{L1}) and the local spacetime will approach a cone spacetime, which is transitive under proper conformal transformations only~\cite{abap}. At this energy, therefore, conformal symmetry naturally becomes the relevant symmetry.

On the other hand, a cosmological constant has already been shown to slow down the propagation of light~\cite{walter}. Considering that, according to de Sitter relativity, high energy photons produce around them a local de Sitter spacetime whose intensity is proportional to the photons energy density, the corresponding cosmological term could act as a geometric refractive index, slowing down their propagation. Since the value of $\Lambda$, and consequently of the refractive index, is larger for higher energy density photons, this effect could provide an explanation for the recently observed delay in high energy gamma--ray flares coming from the center of the galaxy Markarian 501~\cite{magic}. As a matter of fact, this mechanism has already been shown to give a good estimate of the observed delay~\cite{AP0}. Considering that, from this point of view the delay would be a manifestation of quantum gravity, de Sitter relativity could show up as a new paradigm to approach this theory.

\section{Final Remarks}

There are theoretical evidences that ordinary special relativity, whose underlying kinematics is ruled by the Poincar\'e group, breaks down at ultra--high energy densities. When looking for a new special relativity, the most natural generalization is arguably to replace Poincar\'e special relativity by  de Sitter special relativity. This means to assume that, at ultra--high energy densities, the local kinematics is ruled by the de Sitter group. This, in turn, means that any high--energy density process must modify the local structure of spacetime in such a way that the region where the process takes place changes from Minkowski to a de Sitter spacetime.

An important point of de Sitter relativity is that, since the local de Sitter spacetime is essentially a kinematic effect, the source of $\Lambda$ is not the energy--momentum tensor.\footnote{The origin of $\Lambda$ in this theory is actually related to conformal symmetry. In fact, note that de Sitter relativity naturally introduces the conformal generators in the spacetime transitivity. As a consequence, the conformal transformations will naturally be incorporated in the kinematics of spacetime, and the corresponding conformal current~\cite{coleman} will appear as part of the Noether conserved current~\cite{vaxjo}. By studying the consistency of this structure with general relativity, the source of $\Lambda$ is found to be the proper conformal current of matter~\cite{AP0}.} As a consequence, no exotic matter is necessary to explain its existence: any kind of ordinary matter produces it. When applied to the whole universe, a simple estimate based on this theory gives a number not far from the observed value of the cosmological constant. When applied to study the propagation of ultra--high energy photons, it gives a good estimate for the recently observed delay in high energy gamma--ray flares coming from the center of the galaxy Markarian 501. If this delay is a manifestation of quantum gravity, de Sitter relativity can be seen as a new way of approaching the quantum gravity problem.

Let us mention finally that one drawback of the usual models of doubly special relativity is that they are valid only at the energy scales where ordinary special relativity is supposed to break down,\footnote{This restriction is known as the ``soccer-ball problem'' \cite{soccer}.} giving rise to a kind of patchwork relativity. On the other hand, de Sitter relativity is found to be invariant under a simultaneous re-scaling of mass, energy and momentum~\cite{dssr}, and is consequently valid at all energy scales~---~it is a ``universal'' relativity. This is a very important property presented by all fundamental theories, like for example quantum mechanics. We can then say that the above results constitute a compelling indication that the answer to the question in the title is: {\em yes, physics seems to be again asking for a new kinematics}.

\section*{Acknowledgments}
The authors would like to thank FAPESP, CNPq and CAPES for partial financial support.


\end{document}